\UseRawInputEncoding
\documentclass{article}
\usepackage{appendix}
\usepackage[bookmarks=false]{hyperref}
\usepackage{amsmath}
\usepackage{amssymb}
\usepackage[retainorgcmds]{IEEEtrantools}
\usepackage{graphicx}

\begin{document}
\title{Gas in external fields: the weird case of the logarithmic trap}
\author{Loris Ferrari \\ Department of Physics and Astronomy (DIFA) of the University of Bologna\\via Irnerio, 46 - 40126, Bologna,Italy}
\maketitle
\begin{abstract}
The effects of an attractive logarithmic potential $u_0\ln(r/r_0)$ on a gas of $N$ non interacting particles (Bosons or Fermions), in a box of volume $V_D$, are studied in $D=2,\:3$ dimensions. The unconventional behavior of the gas challenges the current notions of thermodynamic limit and size independence. When $V_D$ and $N$ diverge, with finite density $N/V_D<\infty$ and finite trap strength $u_0>0$, the gas collapses in the ground state, independently from the bosonic/fermionic nature of the particles, at \emph{any} temperature. If, instead, $N/V_D\rightarrow0$, there exists a critical temperature $T_c$, such that the gas remains in the ground state at any $T<T_c$, and \textquoteleft evaporates\textquoteright$\:$ above, in a non-equilibrium state of borderless diffusion. For the gas to exhibit a conventional Bose-Einstein condensation (BEC) or a finite Fermi level, the strength $u_0$ must vanish with $V_D\rightarrow\infty$, according to a complicated exponential relationship, as a consequence of the exponentially increasing density of states, specific of the logarithmic trap.  
\end{abstract}

*Corresponding author: e-mail: loris.ferrari@unibo.it

\section{Introduction}
\label{intro}

In recent years, huge technological progresses made it possible to confine neutral particles in a small region of space by magnetic traps \cite{Pr,TCSH,VHZ,FZ,PS}, which results, in the most current cases, in a harmonic potential \cite{Romero}. With the aid of laser cooling tecniques \cite{Ph}, the use of magnetic traps has led to unprecedental results, like the direct observation of BEC \cite{Kal, BST,Frye} and small gravimetric effects \cite{Muntinga,Carraz,Meister, Bravo}. In the mainstream of those results, the theoretical literature on the harmonic traps and on the gravitational field, is well represented \cite{Bagnato,Soldati}.
   
Due to the difficulty of realizing LT's for neutral particles, the gas thermodynamics in a logarithmic trap (LT) has less fields of concrete application. The interest in LT's dates back to ref \cite{Hoov}, which studies the orbits of charged particles in the logarithmic radial field produced by a charged long wire, in the plane perpendicular to the wire itself. A logarithmic baryon-baryon interaction was later suggested as a possible interpretation of the composite hadrons' mass formula \cite{Mur1,Mur2,Paa}. Studies on the gravitational field close to the Schwarzschild radius of black holes \cite{Shak} provide an example of logarithmic potential in a cosmological context. Besides any possible concrete application, the present work primarly aims to stress some surprising aspects of LT's, quite ignored by the current literature, to the author's knowledge. In particular, it will be shown that LT's challenge some standard notions of Thermodynamics, like the thermodynamic limit (TL), the size independence and the semi-classical limit at low gas density and/or high temperatures.

The Hamiltonian:

\begin{equation}
\label{H}
H = \frac{\mathbf{p}^2}{2m}+\begin{cases}
u_0\ln(r/r_0)&\quad (r<R)\\
\\
\infty&\quad (r>R)
\end{cases}
\quad,
\end{equation}
\\
with $u_0>0$ and $r=|\mathbf{r}|$, describes an attractive TL,  in a box of radius $R$ and volume $V_ D=\Omega_DR^D/D$ ($\Omega_D$ being the solid angle in D-dimensions). In Section \ref{DOS} the density of states (DOS) resulting from the Hamiltonian \eqref{H} is calculated in 2 and 3 dimensions, showing the separation between an increasing exponential shape $\mathrm{e}^{D\epsilon/u_0}$, below a characteristic energy $\epsilon_c$, and a free-particle shape $\epsilon^{D/2-1}$ above (Fig. 1). In Section \ref{DOS+CP}, we adopt the current definition of thermodynamic limit (TL):

\begin{equation}
\nonumber
N,\:V_D\rightarrow\infty\quad;\quad \rho:=N/V_D<\infty\:,
\end{equation}
\\
and study the possible relationships between $N$, $V_D$ and the LT's strength $u_0$. In particular, if $V_D$ diverges as follows:  

\begin{center}
(I)
\end{center}
\begin{center}
$V_D\propto\mathrm{e}^{\epsilon_cD/u_0}u_0^{-D/2}\:;\quad u_0\rightarrow0$,
\end{center}
$\:$\\
$\epsilon_c$ being a fixed parameter, independent from $u_0$, then $\epsilon_c$ is shown to act like a gap between the ground state and the free-particle DOS (Subsection \ref{(I)}). In this case, the chemical potential $\mu$ behaves in the current way, showing BEC at a finite temperature, for Bosons, and a finite Fermi level, for Fermions. 

If, instead: 

\begin{center}
(II)
\end{center}
\begin{center}
$V_D\rightarrow\infty,\:u_0>0$,
\end{center}
$\:$\\
one gets a really weird result (Section \ref{(II)simple}): the chemical potential of the bosonic gas tends to the ground state energy $E_0$, and the Fermi level of a fermionic gas tends to diverge, for all $T<\infty$. This means that the gas is frozen in the ground state at \emph{any} temperature, independently from the bosonic or fermionic nature of the particles. Since this forbids any thermal activity of the log-trapped gas, one must assume the presence of a \textquoteleft normal\textquoteright$\:$LT-insensitive gas, which provides the thermal bath at $T>0$. If TL is generalized to include the case of $N$ diverging less rapidly than $V_D$ (Subsection \ref{Off}), a critical temperature $T_c$ can be defined, such that the gas (bosonic or fermionic) is still frozen in the ground state for all $T\le T_c$, but expands itself for $T>T_c$, in a non equilibrium state of borderless diffusion.

In short, the normal behavior of the gas follows from a fairly complicated condition, like (I), involving an exponential divergence of the box size, with the vanishing trap's strength. Instead, what looks the most natural layout (II) - a fixed trap in a divergingly large box - causes a quite unconventional behavior, whose image can be clearly seen in Fig. 2 (Section \ref{EV}), which shows the \emph{effective} volume \cite{PS}, i.e. the mean volume occupied by a single particle, at a given temperature, \emph{modulo} the degeneracy effects and the exclusion principle. EV is shown to increase abruptly from sizes comparable to the volume occupied by the single particle's ground state, to the whole box volume, above the critical temperature $T_D=u_0/D$. It is clearly seen that this is due to the competition between the exponential DOS $\propto\mathrm{e}^{D\epsilon/u_0}$ and the Boltzmann factor $\mathrm{e}^{-\beta\epsilon}$. In Section \ref{Guilty}, we show that the physical origin of the weird results described above is the special nature of the logarithmic potential $u(r)=u_0\ln(r/r_0)$, as a border case between a true trap ($u_+(r)=w_0(r/r_0)^\lambda$) and a potential hole ($u_-(r)=-w_0(r/r_0)^{-\lambda}$). 

For the sake of brevity, the calculations are included in the Supplemental Materials File, and quoted as references in the text.

\section{The density of states (DOS)}
\label{DOS}
\subsection{The continuous limit (CL)}
\label{ACL}

Let $\vec{\alpha}$ be a discrete set of parameters characterizing the excited quantum levels $\epsilon_{\vec{\alpha}}$ and degeneracies $g_{\vec{\alpha}}$ of a generic spin-independent Hamiltonian $h(\mathbf{p},\:\mathbf{r})$. The exact density of states (DOS) in the energy $\epsilon$, measured with respect to the ground level $\epsilon_0=0$, can be written as:
 
 \begin{equation}
 G_{ex}(\epsilon)=g_0\delta(\epsilon)+\overbrace{\sum_{\vec{\alpha}\neq0}g_{\vec{\alpha}}\delta(\epsilon-\epsilon_{\vec{\alpha}})}^{g_{ex}(\epsilon)}
 \end{equation}
 \\
where $\delta(\cdot)$ is the Dirac delta function. The continuous limit (CL) aims to replace the excited-states DOS $g_{ex}(E)$ with a Riemann integrable function:

\begin{subequations}
\label{g,G}
\begin{equation}
\label{g(E)}
g(\epsilon)=
\begin{cases}
\frac{1}{\mathrm{h}^D}\frac{\mathrm{d}}{\mathrm{d}\epsilon}\int_{0<h(\mathbf{p},\:\mathbf{r})<\epsilon}\mathrm{d}\mathbf{p}\mathrm{d}\mathbf{r}&\quad (\epsilon>0)\\
&\quad\quad\quad\quad\quad,\\
0&\quad (\epsilon< 0)
\end{cases}
\end{equation}
\\
such that:

\begin{equation}
\label{G(E)}
G(\epsilon)=(2s+1)\left[g_0\delta(\epsilon)+g(\epsilon)\right]\:,
\end{equation}
\end{subequations}
\\
$s$ being the particle's spin. Including the ground level as a separate term $g_0\delta(\epsilon)$ is a convenient strategy, if the ground state itself can be populated by an arbitrary large number of particles, as is the case of a bosonic gas. When the energy spacings $\Delta_{\vec{\alpha}}$ between nearest neighbor levels tend to vanish in the thermodynamic limit (TL), equations \eqref{g,G} are rigorous, and one speaks about a rigorous continuous limit (RCL). If the $\Delta_{\vec{\alpha}}$'s do not vanish in the TL, and CL follows from a high-temperature approximation $\beta\Delta\epsilon_{\vec{\alpha}}<<1$ for each $\vec{\alpha}$, we speak about an \emph{approximate} continuous limit (ACL), in which the $\Delta\epsilon_{\vec{\alpha}}$'s can be treated as \textquoteleft infinitesimal\textquoteright$\:$ increments, with respect to the thermal energy $\kappa T$. In view of the possible drawbacks inherent to CL, stressed in refs \cite{P, MKH}, especially in 1 and 2D, and in view of applications to finite-sized gases, it makes sense studying possible strategies to improve ACL, for instance, by applying the continuous approximation only to the part of the spectrum whose energy increments are actually smaller than $\kappa T$. If DOS is an increasing function of $\epsilon$, this means re-writing Eq. \eqref{G(E)} as:

\begin{equation}
G(\epsilon)=(2s+1)\left[g_0\delta(\epsilon)+\sum_{\vec{\alpha}\neq0}^{\vec{\alpha}_T}g_{\vec{\alpha}}\delta(\epsilon-\epsilon_{\vec{\alpha}})+g_T(\epsilon)\right]\:,
\end{equation}
\\
where $\epsilon_{\vec{\alpha}_T}$ is the maximum eigenvalue such that $\beta\Delta_{\vec{\alpha}_T}>1$. The Riemann-integrable DOS

\begin{equation}
\label{g*(E)}
g_T(\epsilon)=
\begin{cases}
\frac{1}{\mathrm{h}^D}\frac{\mathrm{d}}{\mathrm{d}\epsilon}\int_{0<h(p,r)-\epsilon_{\vec{\alpha}_T}<\epsilon}\mathrm{d}\mathbf{p}\mathrm{d}\mathbf{r}&\quad (\epsilon>0)\\
&\quad\quad\quad\quad\quad\\
0&\quad (\epsilon< 0)
\end{cases}\:,
\end{equation}
\\
obtained by replacing $h(p,r)$ with $h(p,r)-\epsilon_{\vec{\alpha}_T}$ in Eq. \eqref{g(E)}, becomes thereby temperature-dependent. For the calculation of $\epsilon_{\vec{\alpha}_T}$ in the continuous approximation, we refer to the number of states 

\begin{equation}
\nonumber
N(\epsilon)=g_0+\int_0^\epsilon\mathrm{d}\epsilon'\:g(\epsilon')\:
\end {equation}
\\
with energy smaller than $\epsilon$, and estimate the energy splitting between consecutive levels as the energy difference

\begin{equation}
\nonumber
\Delta\epsilon=\frac{1}{g(\epsilon)}\:,
\end{equation}
\\
corresponding to the increment of a single state. Then the equation

\begin{equation}
\label{epsilonT1}
\kappa T\approx \frac{1}{g(\epsilon_{\vec{\alpha}_T})}
\end{equation}
\\
provides an estimate for $\epsilon_{\vec{\alpha}_T}$. As we shall see in what follows, improving ACL does not have a relevant influence on the results obtained for LT.

\subsection{DOS in a finite-sized LT}
\label{DOSfs}

Let $\phi_0(r)$ be a s-wave, representing the ground state of the LT-Hamiltonian Eq. \eqref{H}. We assume that the localization length $\ell_0$ of $\phi_0(r)$ is small, compared to $R$, i.e., that the rigid-wall confinement has negligible effects on $\phi_0(r)$. Then the Schr\"{o}dinger equation $H\phi_0(r)=E_0\phi_0(r)$ can ignore the lower line of Eq. \eqref{H}, and can be written in the dimensionless form:

\begin{equation}
\label{SchrEq}
[\underbrace{-\nabla_\mathbf{x}+\ln(x)}_{h_{red}}]f(x)=[\overbrace{\frac{E_0}{u_o}+\ln(x_0)}^{A}]f(x)\:.
\end{equation}
\\
by setting $\mathbf{x}:=\mathbf{r}\sqrt{2mu_0}/\hbar$, $\phi_0(r):=f(x)$ and $x_0:=r_0\sqrt{2mu_0}/\hbar$. The LT's ground level is $\epsilon_m=u_0A$. A simple variational calculation \cite{SM-A} yields: :

\begin{subequations}
\label{Defns}
\begin{equation}
\label{Defns1}
 A= 
 \begin{cases}
 0.576\\
 \\
 1.076
 \end{cases}
 \quad;\quad \ell_0=\frac{\hbar}{\sqrt{4u_0m}}\times
 \begin{cases} 
 1&\quad (D=2)\\
&\quad\quad\quad\quad\quad\quad.\\
3&\quad (D=3)
\end{cases}
\end{equation} 
\\
For the next developments, it is useful to introduce the following definitions:

\begin{equation}
\label{Defns2}
r_c:=\frac{\hbar\:\mathrm{e}^{A}}{\sqrt{2mu_0}}\quad;\quad v_c:=\frac{\Omega_Dr_c^D}{D}\quad;\quad\widehat{V}:=\frac{V_D}{v_c}\quad;\quad\epsilon_c:=\frac{u_0}{D}\ln \widehat{V}\:.
\end{equation}
\end{subequations}
\\
Equation \eqref{SchrEq} shows that the length scale $r_0$ in Eq. \eqref{H} simply shifts the energy origin by the quantity $u_0\ln(x_0)$. Hence, it is convenient to define the ground state energy as $E_0=\epsilon_m+u_0\ln x_0$, and the Hamiltonian as: 

\begin{equation}
\label{h}
h(p,\:r) := H(p,\:r)-E_0=\frac{p^2}{2m}+\begin{cases}
u_0\ln(r/r_c)&\quad (r<R)\\
\\
\infty&\quad (r>R)
\end{cases}\quad\quad \:,
\end{equation}
\\
with ground level $\epsilon_0=0$. From Eq. \eqref{Defns1}, the condition $\ell_0<<R$ yields:

\begin{equation}
\label{condu0}
R\sqrt{u_0}>>\frac{\hbar}{\sqrt{4m}} \:.
\end{equation}
\\
For brevity reasons, we take spinless Bosons and Fermions ($s=0$).

The Riemann-integrable part $g(\epsilon)$ of the DOS (Eq. \eqref{g(E)}) can be calculated with the aid of ref \cite{SM-B}, from Eq. \eqref{h}, under the condition \eqref{condu0}:

\begin{subequations}
\label{g(eps)}
\begin{equation}
\label{g(eps1)}
g(\epsilon)=\frac{\widehat{V}\mathrm{e}^{2A}}{4u_0}\times
\begin{cases}
\mathrm{e}^{2(\epsilon-\epsilon_c)/u_0}&\:\:\:\:(\epsilon\le\epsilon_c)\\
\\
1&\:\:\:\:(\epsilon\ge\epsilon_c)
\end{cases}
\quad\quad (D=2)
\end{equation}
\\
\begin{align}
\label{g(eps2)}
g(\epsilon)&=\frac{\widehat{V}\mathrm{e}^{3A}}{u_02\sqrt{27\pi}}\times\nonumber\\
\nonumber\\
&\times
\begin{cases}
\mathrm{e}^{3(\epsilon-\epsilon_c)/u_0}&\:\:\:\:(\epsilon\le\epsilon_c)\\
\\
\mathrm{e}^{3(\epsilon-\epsilon_c)/u_0}\mathrm{Erfc}(\sqrt{3(\epsilon-\epsilon_c)/u_0})+\\
\\
+\frac{2}{\sqrt{\pi}}\sqrt{3(\epsilon-\epsilon_c)/u_0}&\:\:\:\:(\epsilon\ge\epsilon_c)
\end{cases}
\quad\quad (D=3)\:,
\end{align}
\end{subequations}
\\
where use has been made of the definitions \eqref{Defns2}. In Figure 1, the expressions \eqref{g(eps)} are plotted as a function of $\epsilon/\epsilon_c$. It is seen that above $\epsilon_c$, $g(\epsilon)$ tends to (in 3D), or even coincides with (in 2D) the free-particle DOS, and decreases exponentially below. Hence, $\epsilon_c$ determines the energy range above which the kinetic energy's loosing effect prevails over the attractive effect of LT.

\begin{figure}[htbp]
\begin{center}
\includegraphics[width=3in]{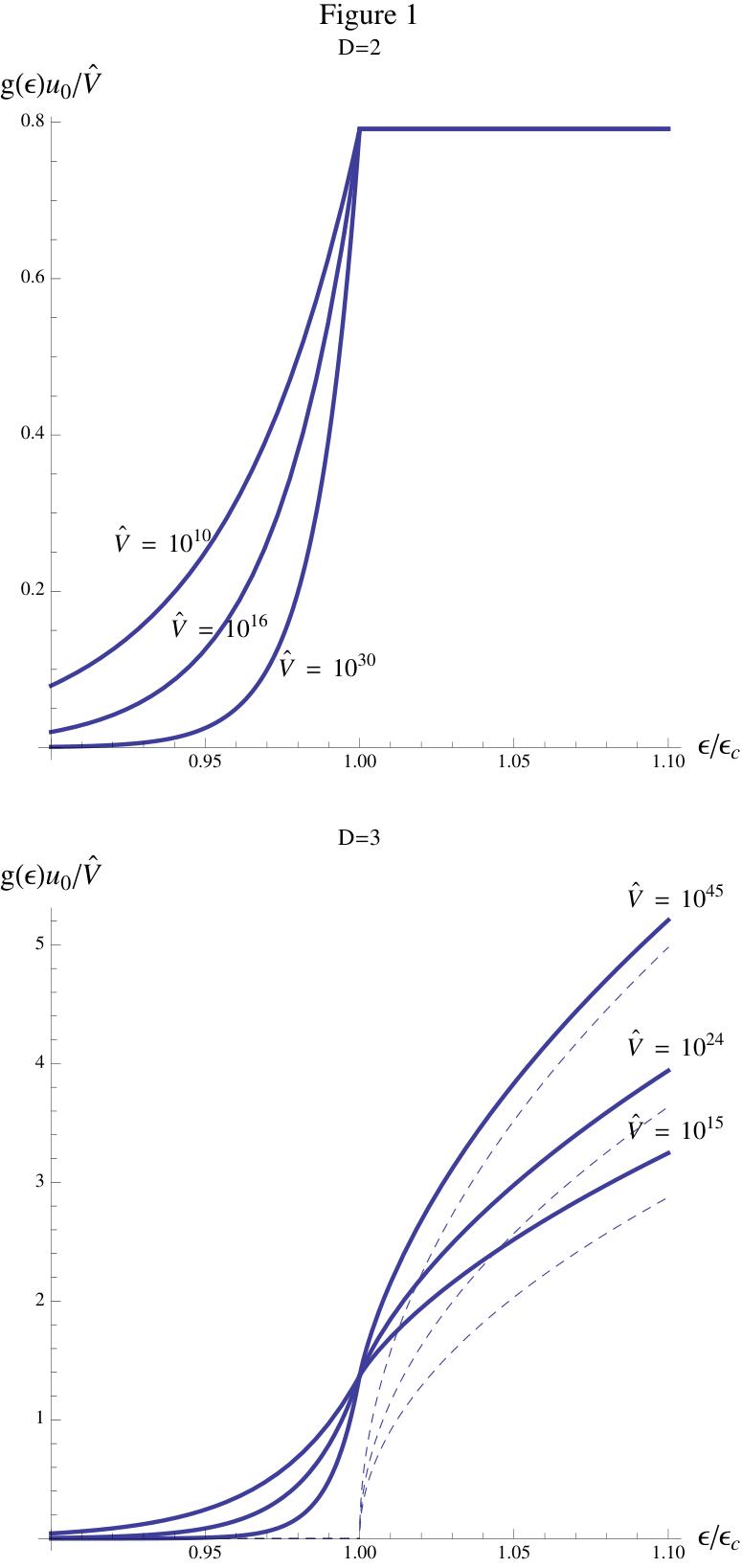}
\caption{\textbf{Riemann-integrable DOS}. Plot of $g(\epsilon)$ in 2 and 3 dimensions, for different values of $\widehat{V}$, as a function of $\epsilon/\epsilon_c$. In $D=2$ $g(\epsilon)$ coincides exactly with the constant free-particle DOS, for $\epsilon>\epsilon_c$. The dashed curves in $D=3$ show the approach to the free-particle shape $\epsilon^{1/2}$, starting from $\epsilon_c$, for $\widehat{V}\rightarrow\infty$. }
\label{default}
\end{center}
\end{figure}

Treating \emph{all} the energy splittings $\Delta\epsilon_{\vec{\alpha}}$ as \textquoteleft infinitesimal\textquoteright$\:$increments in a LT, means taking $\kappa T>>u_0$, because the levels are all proportional to $u_0$. While in case (I) $u_0$ vanishes and RCL does apply, in case (II) $u_0$ is fixed and ACL is to be taken under consideration. Since we are interested in the low energy states at low temperature, we can refer to the case $\epsilon<\epsilon_c$, in Eq.s \eqref{g(eps)}, which involves a pure exponential DOS. A straightforward calculation then yields, from Eq. \eqref{epsilonT1}:

\begin{equation}
\label{epsilonT2}
\epsilon_{\vec{\alpha}_T}\approx \frac{u_0}{D} \ln\left(\frac{u_0}{\lambda_D\kappa T}\right)\quad;\quad\lambda_D=\mathrm{e}^{DA}\times
\begin{cases}
\frac{1}{4}&\:\:\:\:(D=2)\\
\\
\frac{1}{2\sqrt{27\pi}}&\:\:\:\:(D=3)
\end{cases}\:,
\end{equation}
\\ 
for the energy below which the discreteness of the spectrum must be recovered. It is seen that $\epsilon_{\vec{\alpha}_T}$ diverges logarithmically with $T^{-1}$ for vanishing temperature, and becomes \emph{negative} when $T$ exceeds values comparable to

\begin{equation}
\label{TD}
T_D:=\frac{u_0}{D\kappa}\:
\end{equation}
\\
(a special temperature that will come into play soon, in what follows). Since the improved ACL applies to positive eigenvalues $\epsilon_{\vec{\alpha}}$ larger than $\epsilon_{\vec{\alpha}_T}$, equation \eqref{epsilonT2} shows that the improvement is practically uninfluent, for temperatures $T\gtrsim T_D$ and leads to smoothly varying modifications for $T<T_D$.

\subsection{The free-particle limit}
\label{FPlim}

The free-particle limit obviously implies the limit $u_0\rightarrow0$. If the trap is contained in a box, however, the box size must diverge in the same limit, since the localization length of the trap's ground state (and, \emph{a fortiori}, that of the excited states) diverges with vanishing trap's strength (Eq. \eqref{condu0}). In addition, the special case of LT involves the border energy $\epsilon_c(u_0)$, which must vanish in turn. Hence the conditions for the free-particle limit read:

\begin{subequations}
\begin{equation}
\label{condR}
\lim_{u_0\rightarrow0}R\sqrt{u_0}>0 \text{ strictly}\quad,\quad\lim_{u_0\rightarrow0}\epsilon_c= \lim_{u_0\rightarrow0}u_0\ln\left(\frac{R\sqrt{2mu_0}}{\hbar \mathrm{e}^A}\right)=0\:.
\end{equation}
\\
In this case, one is left with the lowest lines of Eq.s \eqref{g(eps)}, which yield: 

\begin{equation}
\label{DOSfree}
\lim_{u_0\rightarrow0}g(\epsilon)=\frac{\Omega_DV_D}{2}\left(\frac{2m}{\mathrm{h}^2}\right)^{D/2}\epsilon^{D/2-1}=g_{free}(\epsilon)\:,
\end{equation}
\end{subequations}
\\
for $\epsilon>0$ (strictly), since $\widehat{V}/u_0^{D/2}$ is proportional to $V_D$.

\section{DOS and chemical potential in the thermodynamic limit (TL)}
\label{DOS+CP}
\subsection{The  case (I)}
\label{(I)}

In contrast to the free-particle limit, in case (I) the border energy $\epsilon_c$ is kept fixed and non zero, for $u_0\rightarrow0$. In this case, on dividing $g(\epsilon)$ by $V_D\propto\widehat{V}/u_0^{D/2}$, it is not difficult to see that the first lines of $g(\epsilon)/V_D$ in Eq.s \eqref{g(eps)} vanish for $u_0\rightarrow0$.\footnote{In particular, the second line in Eq. \eqref{g(eps2)} vanishes because $\mathrm{Erfc}(x)\rightarrow\mathrm{e}^{-x^2}/(x\sqrt{\pi})$ for $x\rightarrow\infty$.} On restoring $V_D$, the lowest lines lead to the expression of the free-particle DOS for $\epsilon>\epsilon_c$, whence:

\begin{equation}
\label{g(eps)I}
\text{(I)}\quad\Rightarrow\quad g(\epsilon)\rightarrow g_I(\epsilon)=
\begin{cases}
0&\quad\quad(\epsilon<\epsilon_c)\\
\\
g_{free}(\epsilon-\epsilon_c)&\quad\quad(\epsilon\ge\epsilon_c)\:,
\end{cases}
\end{equation}
\\
and $G_I(\epsilon)=\delta(\epsilon)+g_I(\epsilon)$ is the total DOS (Eq. \eqref{G(E)}). Figure 1 clearly confirms the content of Eq. \eqref{g(eps)I}, i.e. that $g_I(\epsilon)$ is nothing but the limit of Eq.s \eqref{g(eps)} for $\widehat{V}\rightarrow\infty$, at fixed $\epsilon_c$. Equation \eqref{g(eps)I} shows that LT creates a gap of width $\epsilon_c$, between the ground level $\epsilon_0=0$ and the Riemann integrable DOS, which is simply the free particle DOS, on redefining the energy as $\epsilon'=\epsilon-\epsilon_c$. Hence LT provides a rigorous realization of an approximate model, sometimes used in elementary quantum mechanics, i.e. a potential well, carrying a \emph{single} bound state at the energy $\epsilon'=-\epsilon_c$, whose effect on the unbound states ($\epsilon'>0$) is assumed negligible. We shall return on this point in Section \ref{Guilty}.

Given a generic DOS $G(\epsilon)$, the equation for the chemical potential $\mu$ is:

\begin{equation}
\label{CPeq1} 
1=\int_0^\infty\mathrm{d}\epsilon\frac{G(\epsilon)}{N(\mathrm{e}^{\beta(\epsilon-\mu)}\pm1)}\:,
\end{equation} 
\\
where the double sign refers to Fermions ($+$) and Bosons ($-$). From Eq. \eqref{g(eps)I}, equation \eqref{CPeq1} becomes, with the previously outlined substitution $\epsilon-\epsilon_c\rightarrow\epsilon$:

 \begin{equation}
\label{CP(I)eq1} 
1=\overbrace{\frac{1}{N(\mathrm{e}^{-\beta\mu} \pm1)}}^{n_0}+\underbrace{\int_0^\infty\mathrm{d}\epsilon\frac{\widehat{g}_{free}(\epsilon)}{\mathrm{e}^{\beta(\epsilon+\epsilon_c-\mu)}\pm1}}_{n_\pm(T,\:\mu)}\quad;\quad \widehat{g}_{free}(\epsilon):=g_{free}(\epsilon)/N\:,
\end{equation} 
\\
where $n_0$ is the fractional population of the ground state, and $\widehat{g}_{free}(\epsilon)$ is a size-independent DOS, proportional to $\rho=V_D/N$ (Eq. \eqref{DOSfree}). Of course, $n_\pm(T,\:\mu)$ is the fractional population of the excited states. In the fermionic case ($+$), the quantity of interest is the Fermi level $\epsilon_F=\lim_{T\rightarrow0}\mu$, which can be easily deduced from Eq. \eqref{CP(I)eq1}, since the occupation function $[\mathrm{e}^{\beta(\epsilon-\epsilon_c-\mu)}+1]^{-1}$ becomes a step function for $\beta\rightarrow\infty$ ($T\rightarrow0$), jumping from 1 to 0 at $\epsilon=\epsilon_F-\epsilon_c$. Furthermore, in the TL ($N\rightarrow\infty$) $n_0$ vanishes for any value of $\mu$, whence the equation for $\epsilon_F$ yields:

\begin{equation}
\label{CP(I)2}
1=\int_0^{\epsilon_F-\epsilon_c}\widehat{g}_{free}(\epsilon)\mathrm{d}\epsilon\quad\Rightarrow\quad\epsilon_F=\epsilon_c+\overbrace{\frac{\mathrm{h}^2}{2m}\left(\frac{2\rho D}{\Omega_D}\right)^{2/D}}^{\epsilon_F^{free}}\:,
\end{equation}
\\  
which shows that $\epsilon_F$, in the limit (I), is just the Fermi level $\epsilon_F^{free}$ of the free gas (Eq.\eqref{DOSfree}), plus the border energy $\epsilon_c$.

In the bosonic case ($-$), $\mu>0$ is forbidden by the condition $n_0\ge0$. Until $\mu$ is strictly negative, $n_0$ vanishes in the TL. However, the condition $n_-(T,\:0)<\infty$ for any $T>0$ is necessary and sufficient for the existence of a critical temperature $T_B$ such that $n_-(T_B,\:0)=1$, which leads to (BEC) \cite{Silv,Me2}. From Eq. \eqref{CP(I)eq1} it is easily seen that $\epsilon_c>0$ forbids the vanishing of the denominator in the integrand. Hence $n_-(T,\:\mu)$ is finite for any $T>0,\: \mu\ge0$, and BEC does actually occur at the temperature $T_B=1/(\kappa\beta_B)$, determined by the equation $1=n_-(T_B,\:0)$, which yields:

\begin{equation}
\label{TB(I)}
1=\int_0^\infty\mathrm{d}\epsilon\frac{\widehat{g}_{free}(\epsilon)}{\mathrm{e}^{\beta_B(\epsilon+\epsilon_c)}-1}=\frac{\Omega_D\rho}{2}\left(\frac{2m}{\mathrm{h}^2\beta_B}\right)^{D/2}\overbrace{\int_0^\infty\mathrm{d}x\frac{x^{D/2-1}}{\mathrm{e}^{x+\beta_B\epsilon_c}-1}}^{\mathrm{PolyGamma}\left(\frac{D}{2},\:\mathrm{e}^{-\beta_B\epsilon_c}\right)}\:,
\end{equation}
\\
where $\mathrm{PolyGamma}(\cdot\:,\:\cdot)$ is a special function \cite{Mth}. The resulting $T_B$ increases with $\epsilon_c$, and diverges for $\epsilon_c\rightarrow\infty$. It should be noticed that under condition (I), LT allows BEC even in 2D, which would be impossible, in the free gas case $\epsilon_c=0$. Actually, for $D=2$, $\mathrm{PolyGamma}(1,1)=\infty$.

\subsection{The case (II)}
\label{(II)simple}

Condition (II) corresponds to what looks the simplest possible layout: a fixed LT ($u_0>0$) in a diverging large box ($V_D\rightarrow\infty$). From Eq.s \eqref{g(eps)}, it is immediately seen that this leaves an exponential DOS at any $\epsilon>0$:

\begin{equation}
\label{g(eps)II}
\text{(II)}\quad\Rightarrow\quad g(\epsilon)\rightarrow g_{II}(\epsilon)=
\mathrm{e}^{D\epsilon/u_0}\overbrace{\frac{\mathrm{e}^{DA}}{u_0}\times
\begin{cases}
\frac{1}{4}&\:\:\:\:(D=2)\\
\\
\frac{1}{2\sqrt{27\pi}}&\:\:\:\:(D=3)
\end{cases}}^{g_D}\:.
\end{equation}
\\
From Eq.s \eqref{CPeq1} and \eqref{g(eps)II}, the equation for $\mu$ now reads:

\begin{equation}
\label{eq.musimple} 
1=\frac{1}{N(\mathrm{e}^{-\beta\mu}\pm1)}+\frac{g_D}{N}\int_0^\infty\mathrm{d}\epsilon\frac{\mathrm{e}^{D\epsilon/u_0}}{\mathrm{e}^{\beta(\epsilon-\mu)}\pm1}\:.
\end{equation} 
\\
The competition between the exponential DOS \eqref{g(eps)II} and the Boltzmann factor emerges by noticing that the integral in the r.h.s. of Eq. \eqref{eq.musimple} converges for $T<T_D$ and diverges for $T\ge T_D$ (recall Eq. \eqref{TD}). This makes the solution of Eq. \eqref{eq.musimple} a non trivial task, in the limit $N\rightarrow\infty$. In the Supplemental Material file \cite{SM-C,SM-D}, a rigorous procedure, starting from the finite-sized gas ($\widehat{V},\:N<\infty$) and performing TL after, shows that the equation for the chemical potentials $\mu_-$ (Bosons) and $\mu_+$ (Fermions) can be put in the following form:

\begin{subequations}
\label{zetall}
\begin{align}
\mathrm{e}^{-\beta\mu_\pm}&=\frac{1}{N(1\pm\mathrm{e}^{\beta\mu_\pm})}-\frac{\mathcal{G}T}{T_DN}\ln(1-\mathrm{e}^{\beta\mu_-})\frac{1\mp1}{2}+\\
\nonumber\\
&+
\begin{cases}
\circ(1/N^{\delta_\pm})&(T<T_D)\\
\\
\circ(\ln \widehat{V}/N)&(T=T_D)\\
\\
\frac{\mathcal{G}}{N}\widehat{V}^{1-T_D/T}\mathcal{F}_\pm(T,\:\widehat{V})&(T>T_D)
\end{cases}\label{0zetall}
\end{align}
\end{subequations}
\\
where $\mathcal{G}$ is a numerical factor, $\delta_+=T_D/T$, $\delta_-=1$, and $\mathcal{F}_\pm(\cdot\:,\:\cdot)$ are \emph{strictly bounded} functions of $\widehat{V}$.\footnote{This means that there exists $M(T)<\infty$, such that $\mathcal{F}_\pm(\widehat{V},\:T)<M(T)$ for each $\widehat{V}>0$, $T>T_D$. The factor $(1\mp1)/2$ in Eq. \eqref{zetall} means that the associated logarithmic singularity applies to Bosons only. It is intended that $\circ(x)$ is a quantity proportional to $x$, to leading order.} If $N/\widehat{V}:=\widehat{\rho}$ is finite, it is immediately seen that the expression \eqref{0zetall} vanishes in the TL (in particular, the lowest line vanishes as $V_D^{-T_D/T}$), whence:

\begin{subequations}
\begin{align}
&\mu_-\rightarrow-\frac{\kappa T}{N}\left[1+\frac{\mathcal{G}}{N}\ln(N)\right]\rightarrow0\nonumber\\
&\text{\hspace{6.0cm}for any }T>0\:. \nonumber\\
&\mu_+\rightarrow\infty\nonumber
\end{align}
\end{subequations}
\\
In the bosonic case ($-$), this yields $\mu_-\rightarrow-\kappa T/N$, and the ground state fractional population $n_0$ tends to 1 in the TL. All Bosons are thereby \textquoteleft frozen\textquoteright$\:$in the ground state at \emph{any} temperature. Since the excited states of a Fermi gas involve the occupation of levels lying \emph{above} the chemical potential, the condition $\mu_+\rightarrow\infty$ means that the fermionic gas is frozen in the ground state too, just like the Bosons. 

In conclusion, we get here to the result:
\begin{quote}
\begin{center}
(A)
\end{center}
\emph{Let $N$ Bosons or Fermions be trapped in a LT with fixed strength $u_0>0$, closed in box of volume $V_D$. Then the gas collapses in the ground state at \emph{any} temperature, if $V_D$ diverges, with finite gas density $\rho=N/V_D$ (which is the current form of TL).}
\end{quote}
As mentioned in Section \ref{intro}, a gas in the ground state is thermally inert. Hence, what precedes underlies the presence of what we call a \textquoteleft normal\textquoteright$\:$gas, i.e. a different \emph{LT-insensitive} gas, which provides the thermal bath at $T>0$.

\subsection{From freezing to borderless diffusion: an off-equilibrium transition}
\label{Off}
The competition between the exponential DOS and the Boltzmann factor emerges again from the exponent $1-T_D/T$ in Eq. \eqref{0zetall}, which makes it possible to study an interesting generalization of TL. Let $N=N(\widehat{V})$ be simply a non decreasing function, which warrants the existence of the non negative, but otherwise arbitrary, limit:

\begin{equation}
\label{gamma}
\gamma:=\lim_{\widehat{V}\rightarrow\infty}\frac{\ln N(\widehat{V})}{\ln\widehat{V}}\:.
\end{equation}
\\
The condition $\gamma\le1$ excludes any \textquoteleft overcrowded\textquoteright$\:$ gas ($N/\widehat{V}\rightarrow\infty$), and makes it possible to define the critical temperature:

\begin{equation}
T_c:=\frac{T_D}{1-\gamma}>T_D\:.
\end{equation}
\\
According to Eq. \eqref{gamma}, $N/\widehat{V}$ tends to vanish as $\widehat{V}^{\gamma-1}=\widehat{V}^{-T_D/T_c}$. Hence, equations \eqref{zetall} yield:

\begin{subequations}
\begin{align}
\mathrm{e}^{-\beta\mu_\pm}&=\frac{1}{N(1\pm\mathrm{e}^{\beta\mu_\pm})}-\frac{\mathcal{G}}{N}\ln(1-\mathrm{e}^{\beta\mu_-})\frac{T}{T_D}\left(\frac{1\mp1}{2}\right)+\label{muTc1}\\
&\text{\hspace{8.0cm}}(T>T_D)\nonumber\\
&+\mathcal{G}\left[\widehat{V}^{T_D(1/T_c-1/T)}\mathcal{F}_\pm(\widehat{V},\:T)+\cdots\right]\label{muTc2}
\end{align}
\end{subequations}
\\
showing that the gas is still frozen in the ground state for $T<T_c$, since in this case the $\widehat{V}$-dependent expression in Eq. \eqref{muTc2} vanishes for $\widehat{V}\rightarrow\infty$. In contrast, the same expression \emph{diverges} for $T>T_c$, which yields $\mu_\pm\rightarrow-\infty$. The negative divergence of the chemical potential is an extreme non degeneracy condition which eliminates any difference between the behavior of Fermions and Bosons. In conclusion, the condition $\gamma<1$ and the limit $\widehat{V}\rightarrow\infty$ lead to another weird result:
\begin{quote}
\begin{center}
(B)
\end{center}
\emph{Let $N$ Bosons or Fermions be trapped in a LT with fixed strength $u_0>0$, closed in a box of volume $V_D$. Let $V_D/N$ diverge in the TL. Then there exists a finite critical temperature $T_c$, marking the transition between a total freezing in the ground state below $T_c$, and a state of absolute non degeneracy, above $T_c$.}
\end{quote}

The reason can be understood as follows: if $T<T_c$, the confining effect of the (weak) logarithmic potential is still able to overcome the thermal kinetic energy even in the absence of the rigid-wall confinement. If $T>T_c$, instead, the loosing effect of the kinetic energy prevails, and the lack of a reservoir's confinement makes the particles scatter in an unconfined space, since the number density of particles tends to vanish, due to the condition $\gamma<1$, under which $N$ diverges less rapidly than $\widehat{V}$. This borderless diffusion through the normal LT-insensitive gas (which provides the thermal bath at $T>0$), actually means that the log-trapped gas gets out of thermal equilibrium above $T_c$.

It is important to stress that the results obtained in the limit $\widehat{V}\rightarrow\infty$ are almost insensitive to the improvement of ACL (Subsection \ref{DOSfs}), even for $T$ small compared to $T_D$. Actually, the temperature-dependent DOS Eq. \eqref{g*(E)} is the same as Eq. \eqref{g(eps)}, except for the substitutions:

\begin{equation}
\nonumber
r_c\rightarrow r_c\mathrm{e}^{\epsilon_{\vec{\alpha}_T}/u_0}\quad;\quad \theta_c\rightarrow\theta_c-\frac{D\epsilon_{\vec{\alpha}_T}}{u_0}\quad;\quad \mu\rightarrow\mu-\epsilon_{\vec{\alpha}_T}\:.
\end{equation}
\\ 
Hence, equation \eqref{eq.musimple} becomes:

\begin{subequations}
\label{eq.mu*}
\begin{align}
1&=\frac{1}{N(\mathrm{e}^{-\beta\mu}\pm1)}+\frac{1}{N}\sum_{\vec{\alpha}\neq0}^{\vec{\alpha}_T}\frac{1}{(\mathrm{e}^{-\beta(\mu-\epsilon_{\vec{\alpha}})}\pm1)}+\label{eq.mu*1}\\
\nonumber\\
&+\frac{g_D}{N}\int_0^\infty\mathrm{d}\epsilon\frac{\mathrm{e}^{D\epsilon/u_0}}{\mathrm{e}^{\beta(\epsilon-\mu+\epsilon_{\vec{\alpha}_T})}\pm1}\label{eq.mu*2}
\end{align}
\end{subequations}
\\
The asymptotic behavior of the integral in Eq. \eqref{eq.mu*2}, for $\widehat{V},\:N\rightarrow\infty$ remains the same as in Eq. \eqref{eq.musimple}. The only significant change is that the single contribution of the ground level is replaced by the finite sum in Eq. \eqref{eq.mu*1}, so that in the bosonic case the particles' condensation involves a set of low-energy levels $0\le\epsilon_{\vec{\alpha}}\le\epsilon_{\vec{\alpha}_T}$, instead of the ground level alone. In the fermionic case and for finite $\epsilon_{\vec{\alpha}_T}$, the contribution of the sum simply vanishes in the TL.

\section{The Effective Volume}
\label{EV}
The peculiar features of LT can be clearly illustrated by the behavior of what we call the \emph{effective} volume (EV) \cite{PS,FP}, i.e. the mean volume occupied by a particle at the temperature $T$, \emph{modulo} the degeneracy and the exclusion principle's effects. $V_{eff}$ represents the mean ideal volume available to a particle if it were alone,\footnote{$V_{eff}$ can also be regarded as to the mean volume occupied by a gas of \emph{distinguishable} bosons.} and is useful to illustrate the effects produced by the external fields on the confinement:

\begin{subequations}
\label{Veff,P}
\begin{equation}
\label{Veff}
V_{eff}=\int\mathrm{d}\mathbf{r}\:r_1\cdots r_D\: P(\mathbf{r},\:T)\:,
\end{equation}
\\
where

\begin{equation}
\label{P(r,T)nd}
P(\mathbf{r},\: T)=\frac{|\phi_0(r)|^2+\frac{1}{\mathrm{h}^D}\int_{h(p,\:r)>0}\mathrm{d}\mathbf{p}\:\mathrm{e}^{-\beta h(p,\:r)}}{1+\frac{1}{\mathrm{h}^D}\int_{h(p,\:r)>0}\mathrm{d}\mathbf{p}\:\mathrm{d}\mathbf{r}\:\mathrm{e}^{-\beta h(p,\:r)}}\:,
\end{equation}
\end{subequations}
\\
is the probability distribution of the particle's co-ordinate $\mathbf{r}$, of Cartesian components $r_j$, in the non degeneration limit $\mathrm{e}^{\beta\mu}<<1$. Following the same method as in Section \ref{DOS}, here we introduce the probability density $|\phi_0(r)|^2$ of the ground state as a contribution apart, with respect to the excited states contribution, expressed by the integrals in the r.h.s. of Eq. \eqref{P(r,T)nd}, according to the continuous approximation. In view of TL, a further useful approximation  follows from Fig. 1: as already outlined in Subsection \ref{(I)}, equation \eqref{g(eps)I} can be used as a large $\widehat{V}$ (or $V_D$) approximation, even for the finite-sized gas. In particular, one can approximate the Hamiltonian as a potential well with a single bound state $\phi_0(r)$ at the ground level $\epsilon_0=0$, separated from the free-particle Hamiltonian by a gap of width $\epsilon_c$. In this case the probability density \eqref{P(r,T)nd} reads simply:

\begin{equation}
\label{P(r,T)nd2}
P(\mathbf{r},\: T)=
\begin{cases}
\frac{|\phi_0(r)|^2+\frac{\mathrm{e}^{-\beta\epsilon_c}}{\mathrm{h}^D}\int\mathrm{d}\vec{p}\:\mathrm{e}^{-\beta p^2/(2m)}}{1+\frac{\mathrm{e}^{-\beta\epsilon_c}}{{\mathrm{h}^D}}V_D\int\mathrm{d}\vec{p}\:\mathrm{e}^{-\beta p^2/(2m)}}&\quad\quad (r<R)\\
&\text{\hspace{2.0cm}}.\\
0&\quad\quad (r>R)
\end{cases}
\end{equation}
\\
On expressing the Gaussian integrals and the average volume $v_0$ of the ground state \cite{SM-A} as:

\begin{subequations}
\label{Gauss,v0}
\begin{equation}
\int\mathrm{d}\vec{p}\:\mathrm{e}^{-\beta p^2/(2m)}=\Omega_D(2m/\beta)^{D/2}\mathcal{K}_D\quad,\quad v_0=\left(\frac{\hbar}{\sqrt{2mu_0}}\right)^D\mathcal{C}_D^{-1}
\end{equation}
\\
with:

\begin{equation}
\mathcal{K}_D:=
\begin{cases}
1/2&\\
\\
\sqrt{\pi}/4&
\end{cases},\quad
\frac{1}{\mathcal{C}_D}:=
\begin{cases}
3/2\quad&(D=2)\\
\\
15\sqrt{2}\quad&(D=3)
\end{cases}\:,
\end{equation}
\end{subequations}
\\
equations \eqref{Veff}, \eqref{P(r,T)nd2}, \eqref{Gauss,v0} yield:

\begin{equation}
V_{eff}=\frac{v_0+\frac{\Omega_D^2\mathrm{e}^{-\beta\epsilon_c}}{2\mathrm{h}^DD}R^{2D}(2m/\beta)^{D/2}\mathcal{K}_D}{1+\frac{\Omega_D\mathrm{e}^{-\beta\epsilon_c}}{{\mathrm{h}^D}}V_D(2m/\beta)^{D/2}\mathcal{K}_D}\nonumber\:.
\end{equation}
\\
From the definitions \eqref{Defns2} one has $R=(\widehat{V}/v_c)^{1/D}$, $\exp(\beta\epsilon_c)=\widehat{V}^{T_D/T}$ and $v_c/v_0=\mathrm{e}^{DA}\mathcal{C}_D$, whence the ratio $V_{eff}/v_0$ can be expressed in terms of the dimensionless volume $\widehat{V}=V_D/v_c$ and dimensionless temperature $T/T_D$:

\begin{equation}
\label{ratio}
\frac{V_{eff}}{v_0}=\frac{1+\frac{\mathcal{K}_D\mathcal{C}_DD^{1-D/2}\mathrm{e}^{2DA}}{2(2\pi)^D}(T/T_D)^{D/2}\widehat{V}^{2-T_D/T}}{1+\frac{\mathcal{K}_D\Omega_DD^{1-D/2}\mathrm{e}^{DA}}{(2\pi)^D}(T/T_D)^{D/2}\widehat{V}^{1-T_D/T}}\:.
\end{equation}
\\
\begin{figure}[htbp]
\begin{center}
\includegraphics[width=5.5in]{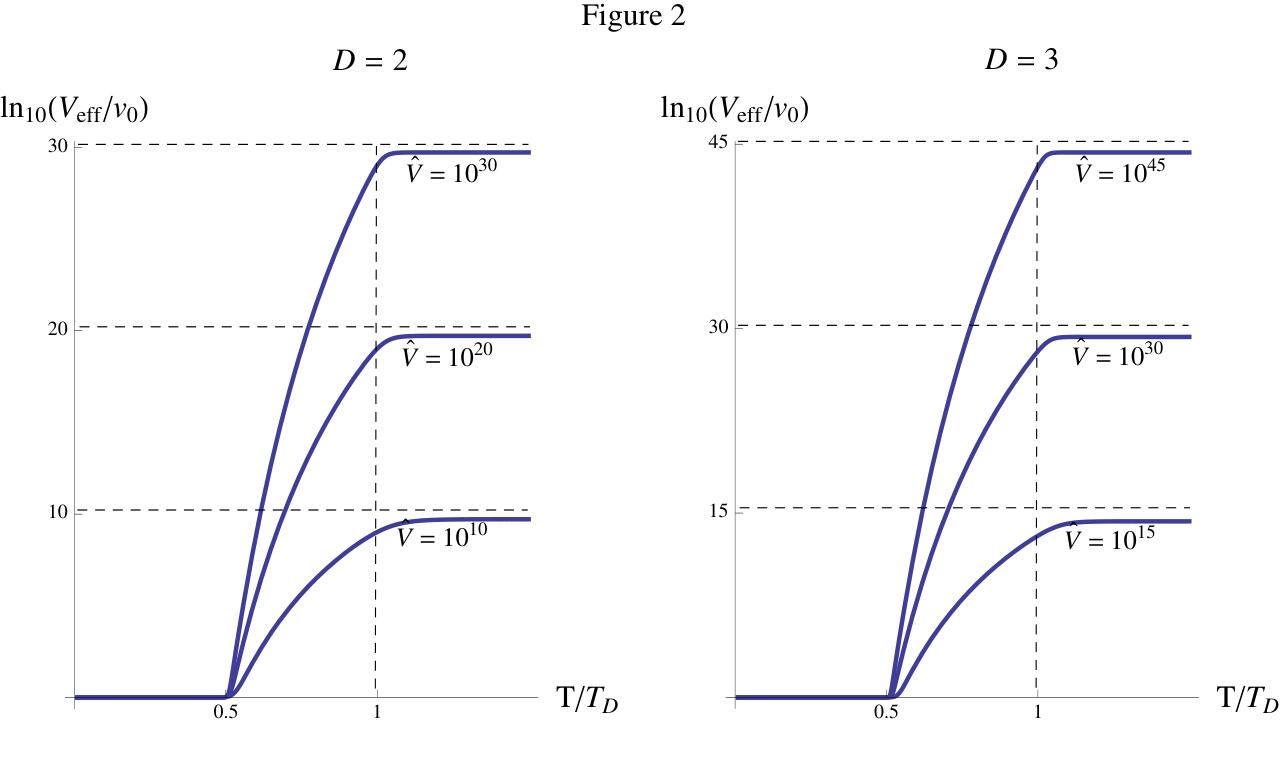}
\caption{\textbf{Effective Volume}. Semi-logarithmic plot of $V_{eff}/v_0$ as a function of $T/T_D$ for different values of $\widehat{V}$, showing the drastic increase of EV for $T>T_D/2$ up to extensive values proportional to $\widehat{V}$ for $T>T_D$. In the logarithmic scale, the differences between $D=2$ and $D=3$ are almost inappreciable.}
\label{default}
\end{center}
\end{figure}

Figure 2 shows a plot of expression \eqref{ratio} for different values of $\widehat{V}$ and $D=2,\:3$. For $T<T_D/2$, EV depends weakly on $T$ and is comparable in magnitude with the minimum size $v_0$, corresponding to the mean volume occupied by the ground state. For $T_D/2<T<T_D$, EV increases rapidly with $T$ and attains an asymptotic value, above $T_D$, proportional to $\widehat{V}$. Equation \eqref{ratio} shows that removing the rigid walls makes EV \textquoteleft explode\textquoteright$\:$ from the mean volume $v_0$ occupied by the single particle in the ground state, to a diverging value, at the critical temperature $T_D$.

\section{Looking for the origin of the LT's weirdness}
\label{Guilty}
The deep origin of the anomalous results expressed by statements (A) and (B) can be found by comparing the DOS Eq.s \eqref{g(eps)} with the DOS's resulting from the Hamiltonians $H_\pm=p^2/(2m)\pm w_0r^{\pm\lambda}$ ($\lambda, w_0>0$), with homogeneous power-law potentials \cite{CEK,Roy,A-H}. $H_+$ yields a \emph{true} trap, which confines the gas even in the absence of any rigid-wall box around.\footnote{A typical example is the harmonic trap with $\lambda=2$. Notice that on setting $w_0=u_0/R^{\lambda}$, the rigid-wall box of radius $R$ is realized by $\lambda\rightarrow\infty$.} The spectrum is positive and discrete up to arbitrary large energies, and TL is realized by the limit $w_0\rightarrow0$, together with CL. In the case $H_-$, instead, the very notion of trap evaporates. Shifting the energy origin to the negative ground level $\epsilon_0$, the spectrum has a discrete branch in the interval $]0,|\epsilon_0|[$, and $|\epsilon_0|$ is an accumulation point of levels $\epsilon_n$, with multiplicity $g_n$, corresponding to bound states with localization length $\ell_n$, such that $\epsilon_n\rightarrow|\epsilon_0|$, $\ell_n\rightarrow\infty$ for $n\rightarrow\infty$.\footnote{A typical example is the Coulombic potential $\lambda=1$.} Above $|\epsilon_0|$, the spectrum is continuous and tends to the free-particle DOS with diverging $\epsilon$. In this case the rigid-wall box is necessary for two related reasons: first, to have the gas confined; second, to have a finite partition function. Actually, the condition that the localization length of the states must be smaller than the box radius ($\ell_n<R$), results in an upper limit $n^*$ for $n$, which makes the diverging series $\sum_{n=0}^\infty g_n\mathrm{e}^{-\beta\epsilon_n}$ become a finite partition function $\sum_{n=0}^{n^*} g_n\mathrm{e}^{-\beta\epsilon_n}$. In short, the passage from $H_+$ to $H_-$ marks the difference between a true trap and what can be loosely called a \textquoteleft potential hole\textquoteright. LT is right the border case between the preceding ones: it is a trap in that the spectrum, \emph{without} the box, is positive and discrete up tu arbitrary large energies; it is a potential hole in that there exists a finite energy $\epsilon_c$ which separates the trapped localized states from those occupying the whole volume \emph{within} the box. However, in contrast to a standard potential hole, in which the border energy $|\epsilon_0|$ depends only on the attractive potential, the border energy 

\begin{equation}
\label{epsc2} 
\epsilon_c=u_0\ln\left(\frac{R\sqrt{2mu_0}\mathrm{e}^{-A}}{\hbar}\right)\:,
\end{equation}
\\
does depend on the box size too, which leads to the difference between the approaches (I) and (II) to TL. As outlined in Subsection \ref{(I)}, the case (I) realizes a perfect potential hole, with a single bound level separated from the free DOS by a gap $\epsilon_c$, held fixed in the TL. The case (II), instead, implements a perfect trap in an infinite space, confining the whole gas in the ground level at any temperature. The origin of the LT's weirdness, is thereby the ambiguous nature of the logarithmic potential, bridging between the true trap's and the potential hole's behavior, which leads to statement (A). The competition between the exponential DOS, specific of LT, and the Boltzmann factor, completes the weirdness, leading to statement (B).

\section{Just a toy model?}
\label{Toy}
 
As a mere toy model, a log-trapped gas is a workbench, aiming to submit some current notions, like TL and size independence, to a sort of stress test, when a logarithmic field is added to the usual rigid-wall confinement in a box. The extreme consequences expressed by statements (A) and (B) show that approaching TL as in case (II) ($u_0>0$, $V_D\rightarrow\infty$) cannot be used as an ideal limit of concrete experimental layouts. However, one could wonder if this would be possible in a finite sized gas, approaching TL as in case (I). As a reference point, let a 3D bosonic gas, with a typical BEC temperature $T_B^0$ of few Kelvin, in the free case, be exposed to the action of a LT, in an appropriate box, such that the relationship \eqref{epsc2}  does yield $\epsilon_c\approx 1 \mathrm{eV}$. Since $\epsilon_c/\kappa T_B^0\approx\mathrm{10}^{4}$, a numerical estimate of the solution of Eq. \eqref{TB(I)} (see Fig. 3) yields $\kappa T_B\approx\epsilon_c/10$. LT would thereby enhance the BEC temperature up to values comparable to $10^3\,^\circ\mathrm{K}$. 

\begin{figure}[htbp]
\begin{center}
\includegraphics[width=4.5in]{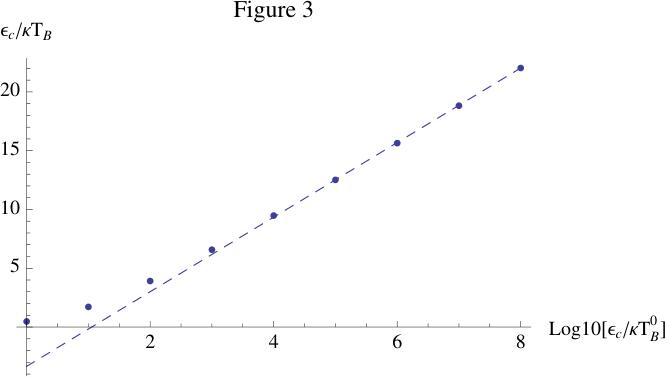}
\caption{\textbf{BEC temperature $T_B$ as a function of $\epsilon_c$ in 3D}. $\epsilon_c/\kappa T_B$ is plotted against the ratio $\epsilon_c/\kappa T_B^0$, $T_B^0$ being the BEC temperature of the free 3D gas. Dots are calculated numerically from Eq. \eqref{TB(I)}. The dashed straight line shows an asympthotic linear interpolation, described by the equation $\epsilon_c/\kappa T_B=3.1713\ln_{10}(\epsilon_c/\kappa T_B^0)-3.3395$.}
\label{default}
\end{center}
\end{figure}

As for Fermions, the Fermi level $\epsilon_F^{free}$ of an atomic free gas in 3D, in standard laboratory conditions ($\rho\approx10^{19}\mathrm{cm}^{-3}$, $m\approx10^{-23}\mathrm{g}$), attains values comparable to $10^{-2}\mathrm{eV}$. With $\epsilon_c\approx1\mathrm{eV}$, equation \eqref{CP(I)2} in Subsection \ref{(I)} would open the possibility of an \emph{atomic} Fermi level  $\epsilon_F=\epsilon_F^{free}+\epsilon_c$ comparable in magnitude with that of the electrons in solids (some $\mathrm{eV}$), which would yield a fermionic \textquoteleft cousin\textquoteright$\:$ of the Bose-Einstein condensate \cite{dMJ}, surviving extremely high temperatures.

What precedes would be astonishingly promising, if it were not for the difficulty of realizing a LT for \emph{neutral} particles. To the author's knowledge, no attempt has been made with magnetic fields. As mentioned in Section \ref{intro}, an attractive logarithmic potential acts on \emph{charged} particles, due to the electric field produced by a very long opposite charged wire, in the plane normal to the wire itself. This, however, calls into play the particle-particle interactions, which can be provisionally ignored (as we did so far) in case of Lennard-Jones potentials between neutral particles, but definitely cannot, in case of Coulombic interactions. So, the possibility of promoting LT from a toy model to a realistic plan for promising laboratory experiences, remains open to future investigations, which are in progress, aiming to study the thermodynamics of \emph{orbitrons} \cite{Hoov,SD-M}, whose core is an electric field realizing a LT for charged particles.

\section{Conclusions}
\label{Conclusions}
A gas in a $D$-dimensional logarithmic trap (LT), confined in a box of volume $V_D$, exhibits very unconventional aspects, which challenge the standard notion of thermodynamic limit (TL), defined as $N,\:V_D\rightarrow\infty$, with $0<N/V_D<\infty$. For a log-trapped gas to exhibit a conventional behavior, characterized by BEC (Bosons), and by a finite Fermi level $E_F$ (Fermions), TL must be accompained by the condition that the energy $\epsilon_c$ (Eq. \eqref{epsc2}) is held fixed at a finite value (case (I)). This implies an exponential divergence of the box radius $R$, with vanishing LT strength $u_0$. If, instead, $u_0>0$ is the fixed parameter (case (II)), the gas (fermionic or bosonic) collapses in the ground state at \emph{any} temperature (statement (A)). This implies the presence of a \textquoteleft normal\textquoteright$\:$ gas insensitive to LT, which provides the thermal bath. If the number of particles $N$ diverges less rapidly than the confinement volume $V_D$, one can define a non negative parameter $\gamma<1$ (Eq. \eqref{gamma}), and a critical temperature $T_c=T_D/(1-\gamma)$, under which the gas is frozen in the ground state, and gets out of thermal equilibrium above, by expanding through the normal gas without limit (statement (B)). Figure 2 provides a clear picture of this off-equilibrium transition, showing the \textquoteleft explosion\textquoteright$\:$of the mean effective volume (EV) occupied by a single particle, \emph{modulo} the degeneration effects. The discussion in Section \ref{Guilty} shows that the origin of the LT's weirdness, is the special nature of the logarithmic potential $u(r)=u_0\ln(r/r_0)$, as a border case, between a true trap ($u(r)=w_0r^{\lambda}$) and a potential hole ($u(r)=-w_0r^{-\lambda}$).

Section \ref{Toy} provides a brief outline of the astonishing consequences resulting from LT's, if those were realizable for \emph{neutral} particles, which looks a very difficult technological task. Studies are in progress to include the Coulombic interactions between the particles, and apply the results to the \emph{ionic} LT's, which are the core of \emph{orbitrons}. 

As a toy model, however, it is possible to use the extreme consequences (A) and (B) to play a mind game, which we call the Logarithmic Universe Game, bridging between the Big Bang \cite{Pee} and the Steady State \cite{Hetal} theory. Ignore any effect of general relativity\footnote{We stress that what follows has nothing to do with the logarithmic potential of ref \cite{Shak}, resulting from general relativity.} and imagine that a LT is placed in the \textquoteleft centre\textquoteright$\:$ of an infinite space ($V_D=\infty$), populated by a \emph{finite} number $N<\infty$ of log-trapped massive particles, at a temperature $T$ provided by a thermal bath of LT-insensitive photons. In this case, one would have $\gamma=\lim_{\widehat{V}\rightarrow\infty}\ln N(\widehat{V})/\ln \widehat{V}=0$ and $T_c=u_0/(D\kappa)=T_D$. For $T<T_D$, all the massive matter would be buried in the ground state. If, however, the photonic gas temperature were raised above $T_D$ (Big Bang), the log-trapped gas would begin to expand almost freely in a just-born LT-universe (obviously non isotropic, due to the LT), according to statement (B). The time scale of diffusion, the possibility of a \textquoteleft dark\textquoteright$\:$matter, still present in the LT ground state, and the continuous creation of visible matter (Steady State), as a result of a still lasting \textquoteleft evaporation\textquoteright$\:$ of the dark matter itself, could all be items for amusing speculations. 

The authors report there are no competing interests to declare

\end{document}


\title{{\Large \textbf{Supplemental material}}\\Gas in external fields: the weird case of the logarithmic trap}
\author{Loris Ferrari \\ Department of Physics and Astronomy (DIFA) of the University of Bologna\\via Irnerio, 46 - 40126, Bologna,Italy}
\maketitle
*Corresponding author: e-mail: loris.ferrari@unibo.it
\\
\\
\begin{appendices}
\numberwithin{equation}{section}

\section{Ground State of the LT}
\label{Appendix A}

When applied to a s-wave $f(x)$, with $x=|\mathbf{x}|$, the Schr\"{o}dinger equation (8) reads
  
\begin{equation}
\label{SchrEq2}
-\left[\partial_x^2+\frac{D-1}{x}\partial_x\right]f(x)+\ln(x)f(x)=Af(x)\:,\\
\end{equation}
\\
in $D=2,\:3$ dimensions. For a qualitative illustration of the LT effects,\footnote{More rigorous treatments can be found in refs \cite{MB,GP,CEK,Roy,A-H}). However, it is seen that the results obtained for the ground-state eigenvalues differ for few percents from those obtained here.} we use the simplest variational form $f(x)=\mathcal{C}\mathrm{e}^{-\lambda x}$ for the ground state solutions, with $\mathcal{C}$ determined by normalization $\int\mathrm{d}^D\mathbf{x}f^2(x)=1$. Then, $\langle\:f\:|h_{red}|\:f\:\rangle$ is minimized with respect to the variational parameter $\lambda$, which yields the elementary result $\lambda=1/\sqrt{2}$ in both cases, and the numerical values reported in Eq. (9a) for $A$ and $\ell_0$ . Note that we estimate the localization length of the ground state as

\begin{equation}
\label{ell0}
\ell_0=\frac{\hbar}{\sqrt{2mu_0}}\int\mathrm{d}^D\mathbf{x}f^2(x)x\:.
\end{equation}
\\
The mean $D$-dimensional volume $v_0$, occupied by the ground state, is given by:

\begin{equation}
\label{v0}
v_0=\left(\frac{\hbar}{\sqrt{2mu_0}}\right)^D\int\mathrm{d}^D\mathbf{x}f^2(x)x^D=\left(\frac{\hbar}{\sqrt{2mu_0}}\right)^D
\begin{cases}
3/2\quad;\quad&(D=2)\\
\\
15\sqrt{2}\quad;\quad&(D=3)
\end{cases}
\end{equation}

\section{The Density of States}
\label{Appendix B}

For the calculation of the DOS Eq.s (11), it is convenient to define the number of states with energy less than $\epsilon$:

\begin{equation}
\label{N}
N(\epsilon):=\frac{1}{\mathrm{h}^D}\int_{0<h(p,\:r)<\epsilon}\mathrm{d}\mathbf{p}\mathrm{d}\mathbf{r}=\frac{(\Omega_D)^2}{\mathrm{h}^D}\int_{0<h(p,\:r)<\epsilon}r^{D-1}\mathrm{d}r\:p^{D-1}\mathrm{d}p\:
\end{equation} 
\\
($\Omega_D=$ solid angle in $D$-dimension) with

\begin{equation}
\nonumber
h(p,\:r) = \frac{p^2}{2m}+\begin{cases}
u(r)&\quad (r<R)\\
\\
\infty&\quad (r>R)
\end{cases}\quad;\quad u(r):=u_0\ln(r/r_c)\:.
\end{equation}
\\
Assuming $R>r_c$ in any case, the condition $0<h<\epsilon$ yields

\begin{equation}
\epsilon-u(r)> \frac{p^2}{2m}>
\begin{cases}
-u(r)\quad&\quad(r<r_c)\\
\nonumber\\
0\quad&\quad(r>r_c)
\end{cases}
\quad\quad (r< R)\:.
\end{equation}
\\
The necessary condition $\epsilon\ge u(r)$ yields, in turn

\begin{equation}
r\le\mathcal{R}(\epsilon):=
\begin{cases}
r_c\mathrm{e}^{\epsilon/u_0}\quad&(\epsilon<u(R))\\
\nonumber\\
R\quad&(\epsilon>u(R))
\end{cases}\:.
\end{equation}
\\
From the preceding definition and formulas, the integral Eq.\eqref{N} becomes:

\begin{align}
&N(\epsilon)=\frac{(\Omega_D)^2(2m)^{D/2}}{D\mathrm{h}^D}\times\nonumber\\
\nonumber\\
&\times\left[\int_0^{\mathcal{R}(\epsilon)}\mathrm{d}r\:r^{D-1}(\epsilon-u(r))^{D/2}-\int_0^{r_c}\mathrm{d}r\:r^{D-1}|u(r)|^{D/2}\right]\:,\nonumber
\end{align}
\\
whose derivation with respect to $\epsilon$ yields:

\begin{align}
g(\epsilon)&=\frac{(\Omega_D)^2(2m)^{D/2}}{2\mathrm{h}^D}\int_0^{\mathcal{R}(\epsilon)}\mathrm{d}r\:r^{D-1}\left[\epsilon-u(r)\right]^{D/2-1}=\nonumber\\
\nonumber\\
&=\frac{(\Omega_D)^2(2m)^{D/2}u_0^{D/2-1}r_c^D}{2\mathrm{h}^D}\mathrm{e}^{D\epsilon/u_0}\underbrace{\int_{\mathcal{L}(\epsilon)}^\infty x^{D/2-1}\mathrm{e}^{-Dx}\mathrm{d}x\label{g1}}_{\mathcal{I_D(\epsilon)}}\:,
\end{align}
\\
where the second line is obtained by the substitution $x=\left[\epsilon-u(r)\right]/u_0$, and

\begin{equation}
\mathcal{L}(\epsilon):=
\begin{cases}
0\quad&(\epsilon<u(R))\\
\nonumber\\
\left[\epsilon-u(R)\right]/u_0\quad&(\epsilon>u(R))
\end{cases}\:.
\end{equation}
\\
The integral in Eq. \eqref{g1} finally reads

\begin{equation}
\mathcal{I}_D(\epsilon)=
\begin{cases}
\mathrm{e}^{-2\mathcal{L}(\epsilon)}/2&\quad(D=2)\\
\nonumber\\
\left[\mathrm{e}^{-3\mathcal{L}(\epsilon)}\sqrt{3\mathcal{L}(\epsilon)}+\frac{\sqrt{\pi}}{2}\mathrm{Erfc}(\sqrt{3\mathcal{L}(\epsilon)})\right]/3^{3/2}&\quad(D=3)\:,
\end{cases}
\end{equation}
\\
which immediately leads to Eq.s (11), when inserted in Eq. \eqref{g1}. The integral $\mathcal{I}_3(\epsilon)$ can be found in the library of Wolfram's \emph{Mathematica} \cite{Mth}.

\section{The Chemical Potential: Bosons}
\label{Appendix C}

Let us rewrite Eq. (21) in the main text as:

\begin{subequations}
\label{eq.mu}
\begin{align}
1=&\frac{1}{N(\mathrm{e}^\zeta\pm1)}+\frac{\mathcal{G}}{N}\overbrace{\int_0^{\theta_c}\mathrm{d}\theta\frac{\mathrm{e}^\theta}{\mathrm{e}^{\sigma\theta+\zeta}\pm1}}^{I_\pm(\theta_c,\:\zeta,\:\sigma)}+\label{I1}\\
\nonumber\\
&+\frac{\mathcal{G}}{N}\mathrm{e}^{\theta_c}
\underbrace{\begin{cases}
\pm\frac{1}{\sigma}\ln[1\pm\mathrm{e}^{-(\sigma\theta_c+\zeta)}]&\quad(D=2)\\
\\
\int_0^\infty\mathrm{d}\theta\frac{\mathrm{e}^\theta\mathrm{Erfc}(\sqrt{\theta})+(2\sqrt{\theta}/\sqrt{\pi})}{\mathrm{e}^{\sigma(\theta+\theta_c)+\zeta}\pm1}\:,&\quad(D=3)
\end{cases}}_{J_\pm(\theta_c,\:\zeta,\:\sigma)} \label{J1}
\end{align}
\\
where:

\begin{equation}
\sigma:=\frac{T_D}{T}\quad;\quad\zeta:=-\beta\mu\quad;\quad\mathcal{G}=\frac{g_Du_0}{D}\:.
\end{equation}
\end{subequations}
\\
In what follows, we shall assume $\theta_c>>1$. It is not difficult to see that $J_\pm(\theta_c,\sigma)$ in Eq. \eqref{J1} is small to order $\mathrm{e}^{-(\sigma\theta_
c+\zeta)}$, for any $\sigma>0$ (i.e. for any finite $T$) and $\zeta\ge0$. In particular:

\begin{equation}
\label{J2}
J_\pm(\theta_c,\:\zeta,\:\sigma)\rightarrow\mathrm{e}^{-(\sigma\theta_
c+\zeta)}\times
\overbrace{\begin{cases}
\frac{1}{\sigma}&(D=2)\\
\\
\frac{1}{\sigma^{3/2}}+\frac{1}{1+\sigma+\sqrt{1+\sigma}}&(D=3)
\end{cases}}^{\mathcal{M}(\sigma)}
\end{equation}
\\
for $\sigma\theta_c>>1$ \cite{Mth}. 

In the bosonic case ($-$), $\zeta$ is non negative\footnote{Otherwise the population of the ground level would be negative.} and $I_-(\theta_c,\:\zeta,\:\sigma)$ diverges for $\zeta\rightarrow0$, hence it is convenient to split the integral as follows:

\begin{equation}
\label{I><}
I_-(\theta_c,\sigma)=\underbrace{\int_0^{1}\mathrm{d}\theta\frac{\mathrm{e}^\theta}{\mathrm{e}^{\sigma\theta+\zeta}-1}}_{I_<(\zeta,\sigma)}+\overbrace{\int_1^{\theta_c}\mathrm{d}\theta\frac{\mathrm{e}^\theta}{\mathrm{e}^{\sigma\theta+\zeta}-1}}^{I_>(\theta_c,\:\zeta,\:\sigma)}
\end{equation}
\\
where the upper integration limit in $I_<$ is set equal to $1$ just for simplicity. After a change of variable ($z=\mathrm{e}^{\sigma\theta}$), an integration by part, a further change of variable ($x=z\mathrm{e}^{-\sigma\theta_c}$), and a suitable rearrangement, one finally extracts the logarithmic singularity:

\begin{equation}
\label{Im}
I_<(\zeta,\:\sigma)=\frac{\mathrm{e}^{-\zeta}}{\sigma}\left[-\ln(1-\mathrm{e}^{-\zeta})+\Im(\sigma,\:\zeta)\right]
\end{equation}
\\
where:

\begin{equation}
\Im(\sigma, \zeta):=\sigma+\mathrm{e}^{(1-\sigma)}\ln(1-\mathrm{e}^{-(\sigma+\zeta)})+\frac{1-\sigma}{\sigma}\int_{\mathrm{e}^{-\sigma}}^1x^{1/\sigma-2}\ln\left(x-\mathrm{e}^{-(\sigma+\zeta)}\right)\mathrm{d}x
\label{Istrange}
\end{equation}
\\
is a bounded function of $\sigma$ and $\zeta$.

If $\sigma>1$ ($T<T_D$), the integral $I_>$ in Eq. \eqref{I><} is regular and converges in the limit $\theta_c\rightarrow\infty$, for any $\zeta\ge0$. In this case equation \eqref{eq.mu} in the TL yields, on account of Eq.s \eqref{Im} and \eqref{J2}:

\begin{align}
1=\frac{1}{N(\mathrm{e}^\zeta\pm1)}-\frac{\mathcal{G}\mathrm{e}^{-\zeta}}{N}\frac{\ln(1-\mathrm{e}^{-\zeta})}{\sigma}+\nonumber&\\
&\quad\quad(\sigma>1)\label{eq.mu<1}\\
\quad\quad+\underbrace{\frac{\mathcal{G}\mathrm{e}^{-\zeta}}{N}\left[\frac{\Im(\sigma)}{\sigma}+\mathrm{e}^\zeta I_>(\infty,\:\zeta,\:\sigma)+\mathcal{M}(\sigma)\right]}_{\circ(1/N)}\:,\nonumber
\end{align}
\\
where $\circ(1/N)$ is a bounded function of $\sigma$ and $\zeta$, which vanishes as $1/N$ for $N\rightarrow\infty$.

In the case $\sigma<1$ ($T> T_D$), $I_>(\theta_c,\:\zeta,\:\sigma)$ diverges for $\theta_c\rightarrow\infty$ and can be expanded in a power series of $\widehat{V}=\mathrm{e}^{\theta_c}$ as follows:

\begin{align}
I_>(\theta_c,\:\zeta,\:\sigma)&=\sum_{n=0}^\infty\mathrm{e}^{-(n+1)\zeta}\int_1^{\ln \widehat{V}}\mathrm{d}\theta\:\mathrm{e}^{-\theta[1-(n+1)\sigma]}=\nonumber\\
\nonumber\\
&=\sum_{n=1}^\infty\frac{\mathrm{e}^{-n\zeta}}{1-n\sigma}\left[\widehat{V}^{1-n\sigma}-\mathrm{e}^{1-n\sigma}\right]\label{I>1}\:,
\end{align}
\\
which yields:

\begin{equation}
\label{I>}
I_>(\theta_c,\:\zeta,\:\sigma)=\mathrm{e}^{-\zeta}\frac{\widehat{V}^{1-\sigma}}{1-\sigma}\left[1+\circ\left(\widehat{V}^{-(1-\sigma)}\right)+\circ\left(\widehat{V}^{-\sigma}\right)\right]\quad\quad(\sigma<1)\:,
\end{equation}
\\
where the higher-order correction is proportional to $\widehat{V}^{-\sigma}$ for $\sigma<1/2$ or to $\widehat{V}^{-(1-\sigma)}$ for $1>\sigma>1/2$.

Taking the limit $\sigma\rightarrow1$ in the first term of the r.h.s. series in Eq. \eqref{I>1}, one finally gets:

\begin{equation}
\label{I>2}
I_>(\theta_c,\:\zeta,\:\sigma)=\mathrm{e}^{-\zeta}\ln \widehat{V}\left[1+\circ(1/\ln \widehat{V})\right]\quad\quad(\sigma=1)\:.
\end{equation}
\\
In conclusion, from Eq.s \eqref{eq.mu<1}, \eqref{I>}, \eqref{I>2}, the equation \eqref{eq.mu} for $\mu$, in the bosonic case reads, to leading order in $\widehat{V}$ and restoring the temperature ($\sigma=T_D/T$) and $\beta\mu=-\zeta$:

\begin{align}
\label{eq.zetabos}
&\mathrm{e}^{-\beta\mu}=\frac{1}{N(1-\mathrm{e}^{\beta\mu})}-\frac{\mathcal{G}T}{T_DN}\ln(1-\mathrm{e}^{\beta\mu})+\\
\nonumber\\
&+
\begin{cases}
\circ(1/N)&\quad\quad(T<T_D)\\
\\
\frac{\mathcal{G}}{N}\ln \widehat{V}\left[1+\circ(1/\ln \widehat{V})\right]&\quad\quad(T=T_D)\:,\\
\\
\frac{\mathcal{G}}{N}\widehat{V}^{1-T_D/T}\mathcal{F}_-(T,\:\widehat{V})&\quad\quad(T>T_D)
\end{cases}
\end{align}
\\
where $\mathcal{F}_-(\cdot,\cdot)$ is a bounded regular function.

\section{The Chemical Potential: Fermions}
\label{Appendix D}
In the fermionic case, the fractional population $n_0$ of the ground level vanishes in the TL, and the integral $I_+(\theta_c,\:\zeta,\:\sigma)$ is bounded for each $\zeta\in\left]-\infty,\:\infty\right[$. So, there is no diverging part to account for. Let:

\begin{equation}
\label{X}
X:=\mathrm{e}^{\sigma\theta_c+\zeta}=\widehat{V}^\sigma\mathrm{e}^{\zeta}\quad,\quad0\le X<\infty\:.
\end{equation}
\\
From Eq.s \eqref{J2} and \eqref{X}, one has:

\begin{align}
\mathrm{e}^{\sigma\theta_c}J_+(\theta_c,\:\zeta,\:\sigma)&=\frac{\mathcal{M}(\sigma)\widehat{V}}{X}\label{J+}\\
\nonumber\\
I_+(\theta_c,\:\zeta,\:\sigma)=\widehat{V}\int_0^{\theta_c}\mathrm{d}\theta\frac{\mathrm{e}^{\theta-\theta_c}}{\mathrm{e}^{\sigma \theta+\zeta}+1}&=\widehat{V}\int_0^{\theta_c}\mathrm{d}z\frac{\mathrm{e}^{-z}}{X\mathrm{e}^{-\sigma z}+1}\:.
\end{align}
\\
The equation for $X$ in the TL ($\theta_c,\:N\rightarrow\infty$) follows from Eq. \eqref{eq.mu}, and reads:

\begin{subequations}
\label{eq.X}
\begin{equation}
\label{eq.X1}
X=\frac{\mathcal{G}\widehat{V}}{N}\mathcal{M}(\sigma)\left[1+\frac{1}{\mathcal{M}(\sigma)}\int_0^{\infty}\mathrm{d}z\frac{\mathrm{e}^{-z}}{\mathrm{e}^{-\sigma z}+1/X}\right]\:,
\end{equation}
\\
which yields:

\begin{equation}
\label{eq.X2}
\frac{\mathcal{G}\widehat{V}}{N}\mathcal{M}(\sigma)<X<\frac{\mathcal{G}\widehat{V}}{N}\mathcal{M}(\sigma)\left[1+\frac{1}{\mathcal{M}(\sigma)(1-\sigma)}\right]\quad\quad(\sigma<1)\:,
\end{equation}
\end{subequations}
\\
since the integral in the r.h.s. of Eq. \eqref{eq.X1} converges to $1/(1-\sigma)$ if $1/X$ is replaced by 0. Recalling the definition \eqref{X}, the inequalities \eqref{eq.X2} show that there exists a \emph{bounded} function $\mathcal{F}_+(\widehat{V},\:\sigma)$ such that:

\begin{align}
&\mathrm{e}^\zeta=\frac{\mathcal{G}\widehat{V}^{1-\sigma}}{N}\mathcal{F}_+(\widehat{V},\:\sigma)\quad\quad(\sigma<1)\nonumber\\
\label{zetaferm<1}\\
\mathcal{M}(\sigma)&<\mathcal{F}_+(\widehat{V},\:\sigma)<\mathcal{M}(\sigma)+(1-\sigma)^{-1}\:.\nonumber
\end{align}
\\
In the case $\sigma>1$ it is convenient to manipulte Eq. \eqref{eq.mu} as follows:

\begin{equation}
\label{eq.zferm}
\mathrm{e}^\zeta=\frac{\mathcal{G}}{N}\left[\widehat{V}^{1-\sigma}\mathcal{M}(\sigma)+\mathrm{e}^\zeta I_+(\infty,\:\zeta,\:\sigma)\right]\:.
\end{equation}
\\
Since for $\sigma>1$ the first term in the r.h.s. brackets vanishes in the TL, one is left with the equation:

\begin{equation}
\frac{N}{\mathcal{G}}=I_+(\infty,\:\zeta,\:\sigma)=\int_0^\infty\mathrm{d}\theta\frac{\mathrm{e}^\theta}{\mathrm{e}^{\sigma\theta-|\zeta|}+1}\:,
\end{equation}
\\
where the \emph{anzatz} that $\zeta$ is negative ($\mu>0$) is to be verified self consistently. Setting $z=\sigma\theta-|\zeta|$, one easily gets \cite{Mth}:

\begin{equation}
\frac{N}{\mathcal{G}}=\frac{\mathrm{e}^{|\zeta|/\sigma}}{\sigma}\int_{-|\zeta|}^\infty\mathrm{d}z\frac{\mathrm{e}^{z/\sigma}}{\mathrm{e}^z+1}=\frac{\mathrm{e}^{|\zeta|/\sigma}}{\sigma}\pi\csc(\pi/\sigma)\left[1+\circ(\mathrm{e}^{|\zeta|(1/\sigma-1)})\right]\:,
\end{equation}
\\
which yields, to leading order:

\begin{equation}
\label{zetaferm>1}
\mathrm{e}^\zeta=\left(\frac{\mathcal{G}\pi\csc(\pi/\sigma)}{N\sigma}\right)^\sigma\left[1+\circ(N^{1-\sigma})\right]\quad\quad(\sigma>1)\:.
\end{equation}
\\
The concluding case $\sigma=1$ is elementary since the integral $I_+$ can be calculated in terms of elementary functions. From Eq. \eqref{eq.zferm}, it is easy to see that:

\begin{equation}
\label{zetaferm=1}
\mathrm{e}^\zeta=\frac{\mathcal{G}}{N}\left[\ln\left(\widehat{V}\frac{\mathrm{e}^\zeta+\widehat{V}^{-1}}{\mathrm{e}^\zeta+1}\right)+\mathcal{M}(\sigma)\right]=\frac{\mathcal{G}}{N}\ln \widehat{V}\left[1+\circ(1/\ln \widehat{V})\right]\quad\quad(\sigma=1)\:.
\end{equation}
\\
In conclusion, the equation for $\zeta$ in the fermionic case yields, according to Eq.s \eqref{zetaferm>1}, \eqref{zetaferm=1}, \eqref{zetaferm<1}:

\begin{equation}
\label{eq.zetaferm}
\mathrm{e}^{-\beta\mu}=\begin{cases}
\left(\frac{\mathcal{G}\pi\csc(\pi T_D/T)T}{T_DN}\right)^{T_D/T}&\quad\quad(T<T_D)\\
\\
\frac{\mathcal{G}}{N}\ln \widehat{V}\left[1+\circ(1/\ln \widehat{V})\right]&\quad\quad(T=T_D)\:.\\
\\
\frac{\mathcal{G}}{N}\widehat{V}^{1-T_D/T}\mathcal{F}_+(T,\:\widehat{V})&\quad\quad(T>T_D)
\end{cases}
\end{equation}
\end{appendices}